\documentclass[aps,prl,superscriptaddress,twocolumn,notitlepage,longbibliography]{revtex4-1}
\usepackage{amsmath}
\usepackage{graphicx}
\usepackage{rotating}
\usepackage{amssymb}
\usepackage{txfonts} 
\usepackage[usenames,dvipsnames]{color}
\usepackage[export]{adjustbox}

\usepackage[colorlinks=true,
            linkcolor=Blue,
            citecolor=Blue,
            urlcolor=Blue,
            pdftitle={Shastry-Sutherland model},
            pdfauthor={A. Keles},
            bookmarks=true
            ]{hyperref}

\usepackage{float}

\usepackage[numbered,open]{bookmark}

\usepackage[all]{hypcap}

\graphicspath{ {./figs/}{../codes/figs/} }

\usepackage{listings}

\usepackage{tikz}

\tikzset{middlearrow/.style={
    decoration={markings,
      mark= at position 0.55 with {\arrow[scale=1,blue]{#1}} ,
    },
    postaction={decorate}
  }
}
\usetikzlibrary{decorations.pathreplacing,
  decorations.markings,decorations.pathmorphing,
    calc,
  shapes}

\usepackage{multirow}

\usepackage{pifont}

\newcommand\x{\ensuremath\mathbf{x}}

\usepackage{bbold}

\newcommand\w{\omega}

\newcommand{\oneloop}[1]
{
\begin{tikzpicture}[scale=.4,baseline=(current bounding box.center)]

    \def \x {0}
    \def \y {.1}
    \def \w {.3}
    \def \l {.5}  
    \def \dver {3}
    \draw[fill=gray] (\x,\y) rectangle (\x+\w,\y+\w);
    \draw[fill=gray] (\x+\w+\dver,\y) rectangle (\x+\dver+2*\w,\y+\w);
    \foreach \m/\n [count=\i] in {#1}
    {
      \ifthenelse{\equal{\i}{1}}
      {
        \ifx\m\n
          \draw[middlearrow={stealth}] 
        \else
          \draw[middlearrow={stealth reversed}] 
        \fi
        (\x-\l,\y+\w+\l) -- (\x,\y+\w);
        \node [left, black] at (\x-\l,\y+\w+\l) 
        {\tiny\ensuremath{\m}};
      }{}  
      \ifthenelse{\equal{\i}{2}}
      {
        \ifx\m\n
        \draw[middlearrow={stealth}] 
        \else
        \draw[middlearrow={stealth reversed}] 
        \fi
        (\x-\l,\y-\l) -- (\x,\y);
        \node [left, black] at (\x-\l,\y-\l) 
        {\tiny\ensuremath{\m}};
      }{} 
      \ifthenelse{\equal{\i}{3}}
      {
        \ifx\m\n
        \draw[middlearrow={stealth}] 
        \else
        \draw[middlearrow={stealth reversed}] 
        \fi
        (\x+2*\w+\dver,\y+\w) -- (\x+2*\w+\dver+\l,\y+\w+\l);
        \node [right, black] at (\x+2*\w+\dver+\l,\y+\w+\l) 
        {\tiny\ensuremath{\m}};
      }{} 
      \ifthenelse{\equal{\i}{4}}
      {
        \ifx\m\n
        \draw[middlearrow={stealth}] 
        \else
        \draw[middlearrow={stealth reversed}] 
        \fi
        (\x+2*\w+\dver,\y) -- (\x+2*\w+\dver+\l,\y-\l);
        \node [right, black] at (\x+2*\w+\dver+\l,\y-\l) 
        {\tiny\ensuremath{\m}};
      }{} 
      \ifthenelse{\equal{\i}{5}}
      {
        \ifx\m\n
        \draw[middlearrow={stealth reversed}] 
        \else
        \draw[middlearrow={stealth }] 
        \fi
        (\x+\w+\dver,\y+\w) 
        to [out=145,in=35]
        (\x+\w,\y+\w);
        \node [above, black] at (\x+\w+.5*\dver,\y+\w+.5)
        {\tiny\ensuremath{\m}};
      }{} 
      \ifthenelse{\equal{\i}{6}}
      {
        \ifx\m\n
        \draw[middlearrow={stealth reversed}] 
        \else
        \draw[middlearrow={stealth }] 
        \fi
        (\x+\w+\dver,\y) 
        to [out=-145,in=-35] 
        (\x+\w,\y);
        \node [below, black] at (\x+\w+.5*\dver,\y-0.5) 
        {\tiny\ensuremath{\m}};
      }{} 
    } 
  \end{tikzpicture}
}

\def\iline(#1,#2,#3,#4,#5,#6)
{
  \draw[
  thick,
  decorate,
  decoration={snake, amplitude=0.3mm, segment length=1.0mm}
  ]
  (#1,#2) -- (#3,#4); 
}
\def\fline(#1,#2,#3,#4,#5,#6)
{
  \draw[middlearrow={stealth}] (#1,#2) to (#3,#4); 
} 
\newcommand{\vertex}[1]
{
\begin{tikzpicture}[scale=.3,baseline=(current bounding box.center)]

    \def \x {0}
    \def \y {.1}
    \def \w {.5}
    \def \l {.6}  
    \def \dver {3}
    \draw[fill=gray] (\x,\y) rectangle (\x+\w,\y+\w);
    \foreach \m/\n [count=\i] in {#1}
    {
      \ifthenelse{\equal{\i}{1}}
      {
        \ifx\m\n
          \draw[middlearrow={stealth}] 
        \else
          \draw[middlearrow={stealth reversed}] 
        \fi
        (\x-\l,\y+\w+\l) -- (\x,\y+\w);
        \node [left, black] at (\x-\l,\y+\w+\l) 
        {\tiny\ensuremath{\m}};
      }{}  
      \ifthenelse{\equal{\i}{2}}
      {
        \ifx\m\n
        \draw[middlearrow={stealth}] 
        \else
        \draw[middlearrow={stealth reversed}] 
        \fi
        (\x-\l,\y-\l) -- (\x,\y);
        \node [left, black] at (\x-\l,\y-\l) 
        {\tiny\ensuremath{\m}};
      }{} 
      \ifthenelse{\equal{\i}{3}}
      {
        \ifx\m\n
        \draw[middlearrow={stealth}] 
        \else
        \draw[middlearrow={stealth reversed}] 
        \fi
        (\x+\w,\y+\w) -- 
        (\x+\w+\l,\y+\w+\l);
        \node [right, black] at (\x+\w+\l,\y+\w+\l)
        {\tiny\ensuremath{\m}};
      }{} 
      \ifthenelse{\equal{\i}{4}}
      {
        \ifx\m\n
        \draw[middlearrow={stealth }] 
        \else
        \draw[middlearrow={stealth reversed}] 
        \fi
        (\x+\w,\y) -- 
        (\x+\w+\l,\y-\l);
        \node [right, black] at (\x+\w+\l,\y-\l) 
        {\tiny\ensuremath{\m}};
      }{} 
    } 
  \end{tikzpicture}
}
\newcommand{\oneloopV}[1]
{
  \begin{tikzpicture}[scale=.8, baseline=(current bounding box.center)]
    \ifthenelse{\equal{1}{#1}}
    {
      \iline(1,0,1,1,d,a )  
      \iline(2,0,2,1,d,a )  

      \fline(0.5,0,1,0,d,a )  
      \fline(1,0,2,0,d,a )  
      \fline(2,0,2.5,0,d,a )  

      \fline(0.5,1,1,1,d,a )  
      \fline(1,1,2,1,d,a )  
      \fline(2,1,2.5,1,d,a )
    }{}  
    \ifthenelse{\equal{2}{#1}}
    {
      \iline(1.5, 0.0, 1.5, 0.3, d, a )  
      \iline(1.5, 0.7, 1.5, 1.0, d, a )  

      \fline(0.5, 0.0, 1.5, 0.0, d, a )  
      \fline(1.5, 0.0, 2.5, 0.0, d, a )  

      \fline(0.5, 1.0, 1.5, 1.0, d, a )  
      \fline(1.5, 1.0, 2.5, 1.0, d, a )  
      \draw[middlearrow={stealth}] (1.5, 0.7) to[out=0,in=0, looseness=1.5]  (1.5,0.3); 
      \draw[middlearrow={stealth}] (1.5, 0.3) to[out=180,in=180, looseness=1.5]  (1.5,0.7);
    }{}  
    \ifthenelse{\equal{3}{#1}}
    {
      \fline(1,0,1.5,.5,d,a )  
      \fline(1.5,.5,2,0,d,a )  
      \iline(1.5,.5,1.5,1,d,a )  

      \fline(0.5,0,1,0,d,a )  
      \iline(1,0,2,0,d,a )  
      \fline(2,0,2.5,0,d,a )  

      \fline(0.5, 1.0, 1.5, 1.0, d, a )  
      \fline(1.5, 1.0, 2.5, 1.0, d, a )
    }{}  
    \ifthenelse{\equal{4}{#1}}
    {
      \fline(1.0, 1.0, 1.5, 0.5, d, a )  
      \fline(1.5, 0.5, 2.0, 1.0, d, a )  
      \iline(1.5, 0.0, 1.5, 0.5, d, a )  

      \iline(1.0, 1.0, 2.0, 1.0, d, a )  
      \fline(0.5, 0.0, 1.5, 0.0, d, a )  
      \fline(1.5, 0.0, 2.5, 0.0, d, a )  

      \fline(0.5, 1.0, 1.0, 1.0, d, a )  
      \fline(2.0, 1.0, 2.5, 1.0, d, a )
    }{}  
    \ifthenelse{\equal{5}{#1}}
    {
      \iline(1,0,2,1,d,a )  
      \iline(2,0,1,1,d,a )  

      \fline(0.5,0,1,0,d,a )  
      \fline(1,0,2,0,d,a )  
      \fline(2,0,2.5,0,d,a )  

      \fline(0.5,1,1,1,d,a )  
      \fline(1,1,2,1,d,a )  
      \fline(2,1,2.5,1,d,a )
    }{}  

  \end{tikzpicture}
}

\newcommand{\siteVertexA}[1]
{
  \begin{tikzpicture}[scale=.8, baseline=(current bounding box.center)]
      \iline(1,0.25,1,.75,d,a )  

      \fline(0.5,0,1,0.25,d,a )  
      \fline(1,0.25,1.5,0,d,a )  

      \fline(0.5,1,1,.75,d,a )  
      \fline(1,0.75,1.5,1,d,a )  
    \node [above, black] at (1,.8) {\tiny\ensuremath{i_1}};
    \node [below, black] at (1,.2) {\tiny\ensuremath{i_2}};
    \node [left, black] at (0.5,1) {\tiny\ensuremath{1}};
    \node [left, black] at (0.5,0) {\tiny\ensuremath{2}};
    \node [right, black] at (1.5,1){\tiny\ensuremath{1'}};
    \node [right, black] at (1.5,0){\tiny\ensuremath{2'}};
  \end{tikzpicture}
}

\newcommand{\siteVertexB}[1]
{
  \begin{tikzpicture}[scale=.8, baseline=(current bounding box.center)]
      \iline(1,0.25,1,.75,d,a )  

      \fline(0.5,0,1,0.25,d,a )  
      \fline(1,0.25,1.5,0,d,a )  

      \fline(0.5,1,1,.75,d,a )  
      \fline(1,0.75,1.5,1,d,a )  
    \node [above, black] at (1,.8) {\tiny\ensuremath{i_1}};
    \node [below, black] at (1,.2) {\tiny\ensuremath{i_2}};
    \node [left, black] at (0.5,1) {\tiny\ensuremath{1}};
    \node [left, black] at (0.5,0) {\tiny\ensuremath{2}};
    \node [right, black] at (1.5,1){\tiny\ensuremath{2'}};
    \node [right, black] at (1.5,0){\tiny\ensuremath{1'}};
  \end{tikzpicture}
}
 
\begin{document} 
\title{Rise and fall of plaquette order in Shastry-Sutherland magnet revealed by 
pseudo-fermion functional renormalization group}
\author{Ahmet Kele\c{s}} 
\email{ahmetkeles99@gmail.com}
\affiliation{Department of Physics, 
  Middle East Technical University, Ankara 06800,
  Turkey }
\author{Erhai Zhao} 
\affiliation{Department of Physics and Astronomy,
  George Mason University, Fairfax, Virginia 22030,
  USA}
\begin{abstract}  
The Shastry-Sutherland model as a canonical example of frustrated magnetism has been extensively studied. The conventional wisdom has been that the transition from the plaquette valence bond order to the Neel order is direct and potentially realizes a deconfined quantum critical point beyond the Ginzburg-Landau paradigm. This scenario however was challenged recently by improved numerics from density matrix renormalization group which offers evidence for a narrow gapless spin liquid between the two phases.  Prompted by this controversy and to shed light on this intricate parameter regime from a fresh perspective, we report high-resolution functional renormalization group analysis of the generalized Shastry-Sutherland model. The flows of over 50 million running couplings provide a detailed picture for the evolution of spin correlations as the frequency/energy scale is dialed from the ultraviolet to the infrared to yield the zero temperature phase diagram.  The singlet dimer phase emerges as a fixed point, the Neel order is characterized by divergence in the vertex function, while the transition into and out of the plaquette order is accompanied by pronounced peaks in the plaquette susceptibility. The plaquette order is suppressed before the onset of the Neel order, lending evidence for a finite spin liquid region for $J_1/J_2\in (0.77,0.82)$, where the flow is continuous without any indication of divergence. 
\end{abstract}
 \maketitle

\begin{figure}[t]
  \centering
  \includegraphics[width=0.49\textwidth]{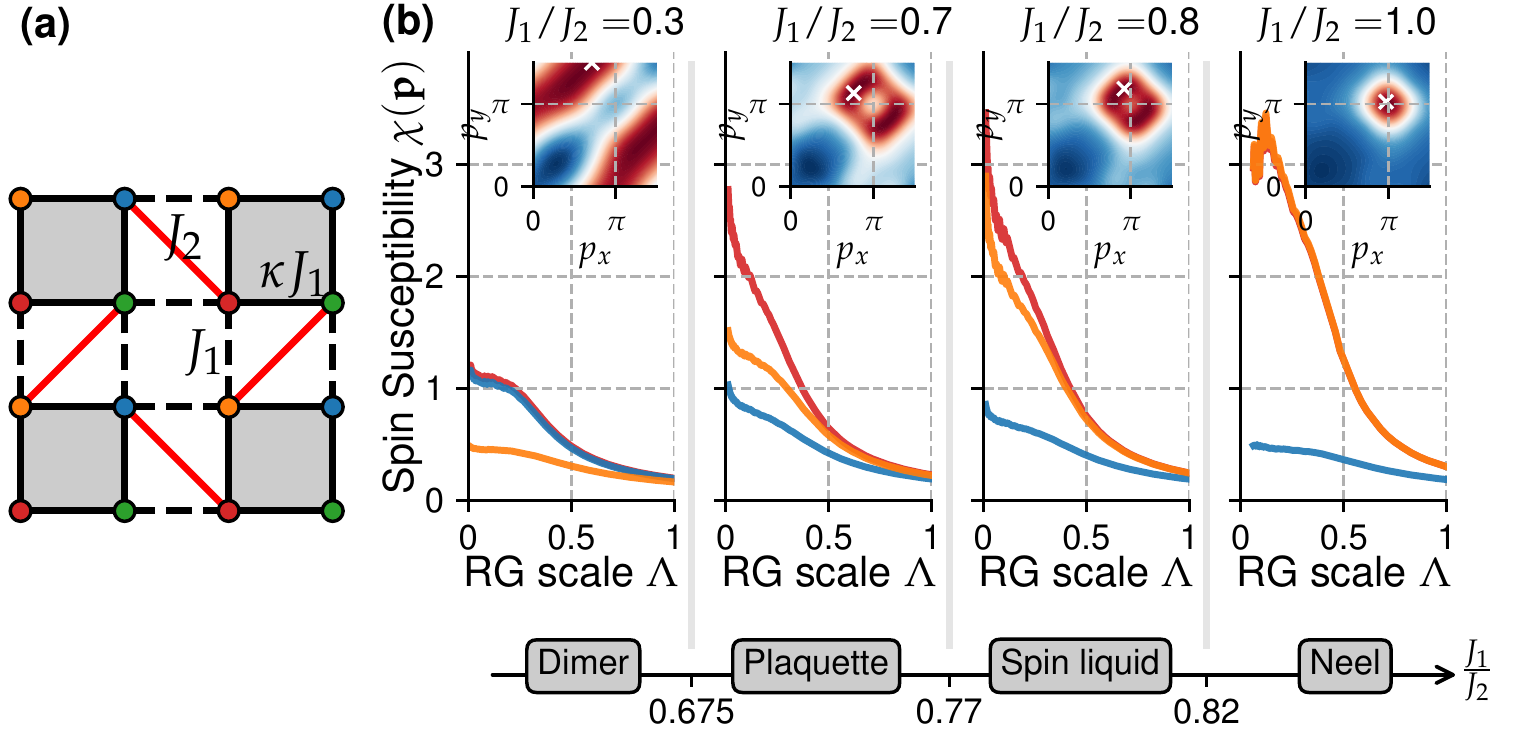}
  \caption{{Generalized Shastry-Sutherland model and its zero temperature
      phase diagram from FRG.} (a) The competing exchange couplings in model
    Eq. \eqref{eq:ss_model} include $\kappa J_1$ and $J_1$ for the black solid
    and dashed bonds respectively, and $J_2$ for the red bonds. The four sites
    within the unit cell (shaded square) are labelled by color red, green,
    blue and yellow respectively. (b) The phase diagram for $\kappa=1$
    consists of four phases separated by three critical points as $J_1/J_2$ is
    varied. A representative point is chosen for each phase to illustrate the
    typical RG flows of the spin susceptibilities $\chi(\mathbf{p})$ in
    leading channels (i.e. different values of $\mathbf{p}$, see main text).
    The corresponding insets show the profile of
    $\tilde{\chi}_{r}(\mathbf{p})$ in the infrared limit $\Lambda\rightarrow
  0$.  The white $\times$ indicates its peak position in the momentum space
  $(p_x,p_y)$. 
    }
  \label{fig:model}
\end{figure}

Forty years after the introduction of the Shastry-Sutherland (SS) model
\cite{SRIRAMSHASTRY19811069}, its ground state phase diagram remains
inconclusive. The model describes quantum spins  on the square lattice with
competing antiferromagnetic exchange interactions, $J_1$ for the
horizontal/vertical bonds and $J_2$ for the decimated diagonal bonds
connecting the empty plaquettes, see Fig.~\ref{fig:model}. Owing to the
frustration, the
model has long been suspected to host exotic ground states and phase
transitions.  A large body of theoretical works have established the existence
of three phases, see e.g. \cite{Miyahara_2003} and \cite{PhysRevX.9.041037,yang2021quantum}
for a synopsis of earlier and recent results, respectively.  The $J_1< J_2/2$ limit
is exactly solvable and the ground state is a product state of diagonal dimers
(spin singlets). For intermediate value of $J_1/J_2$, the ground state is a
plaquette valence bond solid, while Neel order takes over for large $J_1/J_2$.
The most interesting, and controversial, question regards the nature of the
plaquette-to-Neel (pN) transition: is it conventional, a deconfined quantum
critical point, or through an additional spin liquid phase?

Remarkably, the SS model has an almost ideal realization in
$\mathrm{SrCu_2(BO_3)_2}$ crystals, where phase transitions can be induced by
tuning the hydrostatic pressure
\cite{PhysRevLett.82.3168,PhysRevLett.82.3701}.  Inelastic neutron scattering
found signatures of the plaquette phase \cite{zayed20174}, and heat capacity
measurements confirmed the dimer-to-plaquette transition
\cite{PhysRevLett.124.206602,Jimnez2021}.  Yet a direct plaquette-to-Neel
transition was not observed in the anticipated pressure range.  These
experiments renewed the effort to examine this intriguing region using  the
state-of-the-art numerical techniques. Earlier tensor network (iPEPS)
calculations confirmed the plaquette phase within the region
$J_1/J_2\in[0.675,0.765]$
\cite{PhysRevLett.84.4461,PhysRevB.87.115144,PhysRevB.100.140413,PhysRevResearch.1.033038,shi2021phase}
and a weak first order pN transition.
A recent density matrix renormalization group (DMRG) study
\cite{PhysRevX.9.041037} with cylinders of circumference up to ten sites
yielded similar phase boundary $J_1/J_2\in[0.675,0.77]$ but a continuous pN
transition with spin correlations supporting a deconfined quantum critical
point. Another DMRG using cylinder circumference up to 14 sites concludes that
a spin liquid phase exists in the window $J_1/J_2\in[0.79,0.82]$ between the
plaquette and the Neel phase \cite{yang2021quantum}. A core difficulty in
reaching a consensus is attributed to the near degeneracy of the competing orders in this
region. The finite-size limitation of DMRG means that the ground state can
only be inferred by extrapolation via careful finite size scaling analysis.

The size restriction prompts us to adopt an alternative approach diametrically
opposed to exact diagonalization or DMRG on finite systems.  The
algorithm  directly accesses the infrared and thermodynamic limit while
treating all competing orders on equal footing without bias. It starts from
the microscopic spin Hamiltonian and successively integrates out the higher
frequency fluctuations with full spatial (or equivalently momentum) resolution
retained at each step. The scale-dependent effective couplings and correlation
functions are obtained by numerically solving the Functional Renormalization
Group (FRG) flow equations \cite{POLCHINSKI1984,WETTERICH199390,morris1994}. As the frequency scale $\Lambda$ slides from
$J_{1,2}$ down to zero, the zero temperature phase diagram is determined. Such
FRG approach to quantum spin systems, first established in 2010
\cite{PhysRevB.81.144410}, has yielded insights for many frustrated spin
models.
But its application to the SS model has not been successful, perhaps
due to two reasons. First, in contrast to the Neel order, the dimer or the
plaquette order cannot be inferred naively from the divergence of vertex
functions, making it challenging to locate their phase boundaries.  Second, as
we shall show below, the pN transition region is better understood by
examining a generalized model that reduces to the SS model in a particular
limit.

In this work, high resolution FRG analysis of the generalized SS model is
achieved by overcoming these technical barriers.  To maintain sufficient momentum
and frequency resolution, one must keep track of millions of running couplings at
each FRG step.  The calculation is made possible by migrating to the GPU
platform which led to performance improvement by orders of magnitude \cite{PhysRevLett.120.187202,PhysRevB.97.245105}.  Despite
being a completely different approach, the phase boundaries predicted from our
FRG are remarkably close to the state-of-the-art DMRG. The agreement further
establishes FRG as an accurate technique for frustrated quantum
magnetism.  Most importantly, the plaquette susceptibility from FRG indicates
the plaquette order terminates around $J_1/J_2\approx 0.77$ before the onset
of weak Neel order around $J_1/J_2\approx 0.82$. It supports the existence of
a spin liquid region between the plaquette and Neel phase proposed in Ref.
\cite{yang2021quantum}.
Thus, the SS model is a strong
candidate to host spin liquid, and $\mathrm{SrCu_2(BO_3)_2}$ offers exciting
opportunity to realize and probe the elusive spin liquid phase. 


\emph{Model and pseudofermion FRG.---}
Our starting point is the generalized Shastry-Sutherland Hamiltonian \cite{PhysRevB.66.014401}
\begin{equation}
H = \kappa J_1 
    \sum_{\langle i, j\rangle'} \mathbf{S}_i\cdot \mathbf{S}_j 
    + J_1\sum_{\langle i, j\rangle''} \mathbf{S}_i\cdot \mathbf{S}_j 
+ J_2 \sum_{i,j\in \text{diag}} \mathbf{S}_i\cdot \mathbf{S}_j 
\label{eq:ss_model}
\end{equation}
where $\mathbf{S}_i$ are spin one-half operators ($S=1/2$), $i,j$ label the
sites, and $J_{1,2}>0$ are antiferromagnetic exchange couplings.  The first
(second) sum is over nearest neighbors on the square lattice represented by
the solid (dashed) black lines in Fig.~\ref{fig:model}(a), the last sum is over
the alternating diagonal bonds indicated by the red lines.  The original SS
model corresponds to the limit $\kappa=1$ \cite{SRIRAMSHASTRY19811069,Miyahara_2003}. A small 
$\delta J_1=(\kappa-1)J_1$ acts as a source field to break the double degeneracy and
favor valence bond order within the shaded plaquettes. It plays a crucial role
in our analysis and facilitates the calculation of plaquette susceptibility.
We will stay close to the limit $\kappa\rightarrow 1$ throughout.

To predict the phase diagram of Hamiltonian Eq.~\eqref{eq:ss_model}, FRG finds
its generating functional, i.e. an effective field theory 
parametrized by self-energies, four-point and higher-order vertices, 
for each given frequency/energy scale $\Lambda$.
The self-energies and vertices obey the formally exact flow equations that
can be truncated and solved numerically. More specifically,
the many-spin problem is first converted to an interacting fermion problem 
using the pseudofermion representation \cite{PhysRevB.81.144410},
$S_i^\mu=(1/2)\sigma^\mu_{\alpha\beta}\psi^\dagger_{i\alpha}\psi_{i\beta}$.
Here $\sigma^\mu$ are the Pauli matrices, and $\psi_{i\beta}$ annihilates a fermion
at site $i$ with spin $\beta=\uparrow, \downarrow$ etc.. The resultant fermion
Hamiltonian only has quartic interactions but no kinetic energy term (the fermions
are localized and constrained at one particle per site).
So the bare single-particle
Green function $G_0(\omega)=1/i\omega$ with $\omega$ being the
frequency \cite{PhysRevB.81.144410}. Then the flow equations
for the interacting fermion problem can be solved by generalizing the
expansion and truncation schemes extensively benchmarked for 
strongly correlated electronic materials
\cite{RevModPhys.84.299,kopietz2010introduction}. 

The implementation 
of psuedofermion FRG are well documented in the original work
\cite{PhysRevB.81.144410} and later improvements
\cite{
  PhysRevB.94.235138,
  PhysRevB.94.140408,
  PhysRevB.94.224403,
  PhysRevB.95.054418,
  PhysRevB.96.045144,
  PhysRevB.97.064415,
  PhysRevLett.120.057201,
  PhysRevB.99.100405,
  PhysRevB.100.125164,
  PhysRevX.9.011005,
  PhysRevB.103.184407,
  PhysRevB.103.104431, 
  kiese2021multiloop,
  hering2021dimerization}. 
A brief outline is as follows.  Starting from an ultraviolet scale
$\Lambda=\Lambda_{UV}\gg J_{1,2}$ and using the bare interaction in Eq.
\eqref{eq:ss_model} and bare Green function to set up the initial condition,
the coupled integro-differential equations for the scale-dependent
self-energy $\Sigma^\Lambda(\omega)$ and four-point vertex
$\Gamma^\Lambda_{i_1i_2}(\omega_1',\omega_2';\omega_1,\omega_2)$ are solved
successively in small steps along a discretized grid of the frequency/energy
scale $\Lambda$ until it is reduced down to the infrared
$\Lambda=\Lambda_{IR}\rightarrow 0$.  During the flow, the self-energy
$\Sigma^\Lambda(\omega)$ is renormalized to gain nontrivial frequency
dependence as higher frequency fluctuations induce retarded interactions. But
it remains site-independent, i.e. fermions hopping is prohibited.
The four-point vertices
$\Gamma^\Lambda$ (effective interactions) carry multiple indices:  $i_1$
and $i_2$ for lattice sites whereas $\omega_1',\omega_2'$ and
$\omega_1,\omega_2$ are frequencies for the pair of sites before and after the
interaction. Contributions from higher order vertices are approximated by the
Katanin term \cite{PhysRevB.70.115109}.

Care must be exercised to efficiently parametrize the vertices in
order to render the numerical task tractable. In particular, the SS model has
non-symmorphic lattice symmetry, with four sites per unit cell shown in
colors $\alpha=r, g, b, y$ in Fig.~\ref{fig:model} and no $C_4$ symmetry as in
the $J_1$-$J_2$ model.  To avoid using color indices in $\Gamma^\Lambda_{i_1i_2}$,
we pick a $\alpha=r$ site located at the origin as $i_1$. Other vertices for
sites of different color $\alpha=g,b,y$ can be obtained from the central
$r$-site with appropriate rotation and lattice translation
\cite{supplementary}. 
We retain all $i_2$ within a radius
$|\mathbf{r}_{i_1}-\mathbf{r}_{i_2}|<R_\mathrm{max}$ in
$\Gamma^\Lambda_{i_1i_2}$ and emphasize that 
the FRG equations describe infinite systems without a boundary. Here $R_{max}$
merely places an upper cutoff for the correlations retained in the numerics.
As to the frequency variables, we rewrite
$\Gamma^\Lambda(\omega_1',\omega_2';\omega_1,\omega_2)$ as functions of the
Mandelstam variables $s$, $t$ and $u$ \cite{PhysRevB.81.144410} which
manifestly enforce the frequency conservation. Finally, we discretize the
frequency using a logarithmic mesh of $N_\omega$ points
extending from the ultraviolet scale $\Lambda_{UV}=10^2J_2$ to the 
infrared scale $\Lambda_{IR}=10^{-2}J_2$. Typically,
$N_\omega=48$ provides good frequency resolution, and further increasing
$N_\omega$ will not alter the results appreciably. We take $R_{max}=10$ which
amounts to $N_L=441$ lattice sites within the correlation radius. In total,
this gives a coupled system of $N_L\times N_\omega^3\sim$ 50 million running
couplings. 


\emph{Correlation functions and susceptibilities.---} To detect the emergence
of long range order as $\Lambda\rightarrow 0$, correlation functions at each
renormalization scale can be obtained from the $\Sigma^\Lambda$ and
$\Gamma^\Lambda$ via standard calculations involving Feynman diagrams. For
example, the spin-spin correlation function is given by 
\begin{eqnarray} 
  \chi_{ij}(\omega) &=& \int_0^\infty d\tau 
    e^{i\omega\tau} 
  \langle T S_{i}^z(\tau) S_{j}^z(0) \rangle 
  \nonumber \\ 
  &=& 
  \begin{tikzpicture}[scale=.5,
    baseline={([yshift=-.1cm]current bounding box.center)}]
    
    \coordinate (p1) at (1,1); 
    \coordinate (p2) at (3,1); 
    \coordinate (p3) at (6,1); 
    \coordinate (p4) at (10.4,1); 
      \draw[fill=black] (p1) circle (3pt) node[anchor=east] {$S_i^z$};
      \draw[fill=black] (p2) circle (3pt) node[anchor=west] {$S_j^z$};
      \draw[fill=black] (p3) circle (3pt) node[anchor=east] {$S_j^z$};
      \draw[fill=black] (p4) circle (3pt) node[anchor=west] {$S_j^z$};
      
      \draw[middlearrow={stealth}] 
      (p1) to [out=75,in=105] (p2) ;
      \draw[middlearrow={stealth reversed}] 
      (p1) to [out=-75,in=-105] (p2) ;

      \node[black] at (4.5,1) {$+$};

      \draw[middlearrow={stealth}] 
      (p3) to [out=85,in=125] (8.0,1.15) ;
      \draw[middlearrow={stealth reversed}] 
      (p3) to [out=-85,in=-125] (8.0,0.75) ;
    
      \draw[fill=gray] (8,0.75) rectangle (8.4,1.15);
     
      \draw[middlearrow={stealth}] 
      (8.4,1.15) to [out=55,in=95] (p4) ;
      \draw[middlearrow={stealth reversed}] 
      (8.4,0.75) to [out=-55,in=-95] (p4) ;
      \label{szsz} 
    \end{tikzpicture}, 
\end{eqnarray}
where black dots represent the spin matrix $S_i^z=\sigma^z/2$, filled square
represents  vertex $\Gamma^\Lambda$ and lines with arrows are dressed Green
functions that contain the self-energy.  The scale dependence of $\chi$ is
suppressed for brevity.  We find that it is necessary to distinguish the flows
of spin correlations for different bonds, i.e. pairs of $(i,j)$, because the
symmetry breaking patterns in the SS model are rather complex and involving
valence bond orders. For a given site $i$ of color $\alpha$, one can find
$\tilde{\chi}_\alpha(\mathbf{p})$, the Fourier transform of Eq. \eqref{szsz}
in the limit of $\omega\rightarrow 0$.  It is also convenient to define spin
susceptibility
\begin{equation}
 \chi(\mathbf{p}) = \lim_{\omega\rightarrow 0}\frac{1}{4} \sum_{\alpha} \sum_j
    e^{i\mathbf{p}\cdot(\mathbf{r}_{\alpha}-\mathbf{r}_j)}
    \chi_{ij}(\omega) = \frac{1}{4} \sum_{\alpha}   \tilde{\chi}_\alpha(\mathbf{p}),  
    \label{eq:susceptibility}
\end{equation}
where the $\alpha$ sum is over the four sites of different colors within the
unit cell, the $j$ sum is over all sites, and the limit $\omega\rightarrow 0$
is taken in the end.  The spin susceptibility defined in Eq.
\eqref{eq:susceptibility} has no bond resolution, but its divergence (or lack
thereof) and its profile in momentum space offer a quick diagnosis of the
incipient orders as the ratio $J_1/J_2$ is changed.
Finally, we define a set of plaquette susceptibilities to detect the plaquette
valence bond order.  They measure the bond-resolved responses, e.g. the change
in $\chi_{ij}$, due to a small bond modulation 
\begin{equation}
 {\chi}^P_{ij} \equiv \left. -\frac{1}{J_1} \frac{\partial \chi_{ij}}{\partial \kappa}  \right|_{\kappa\rightarrow 1}
 =  \left. -\frac{\partial \chi_{ij}}{\partial (\delta J_1)}\right|_{\delta J_1\rightarrow 0}
 \label{eq:plaquette-susceptibility}
\end{equation}
with $J_1$ and $J_2$ fixed.
A dramatic enhancement of the $\chi_{ij}^P$ around the shaded plaquettes
indicates an instability against a small fluctuation of modulation $\delta J_1$.
To compute ${\chi}_{ij}^P$, we perform two runs of FRG flow with bare couplings
$(\kappa J_1=J_1+\delta J_1,
J_1,J_2)$ and $(J_1,J_1, J_2)$ for a given bond $(i,j)$.
The procedure is expensive but
provides invaluable insights.


\emph{Phase diagram.---} The final results of our FRG calculation are
summarized in the 
phase diagram shown in Fig.
\ref{fig:model}(b). It contains four phases as $J_1/J_2$ is varied at fixed
$\kappa=1$.  For each phase, a representative value of $J_1/J_2$ is chosen to
illustrate its characteristic FRG flow pattern in two complementary ways.
First is the momentum distribution 
$\tilde{\chi}_{r}(\mathbf{p})$ near the
end of the flow (insets), where the peak momenta are labelled by white
`$\times$' in the extended Brillouin zone \footnote{Typically
  ${\chi}(\mathbf{p})$ is plotted, which is obtained by summing over
  $\tilde{\chi}_{\alpha}$ according to Eq. \eqref{eq:susceptibility} to obey
  all crystal symmetries within the Brillouin zone. Here we choose to show
  $\tilde{\chi}_{r}$ because it appears less cluttered.}.  Next is the flow of
spin susceptibility ${\chi}(\mathbf{p})$ with the RG scale $\Lambda$ (main
panels) for different channels, i.e., different values of $\mathbf{p}$. For
example, the channel $\mathbf{p}=(\pi,0)$ is shown in blue, the $(\pi,\pi)$
channel in orange, while the flow for the peak momenta labelled by  `$\times$'
is in red. Clearly, {\it the leading channels for the four phases are
  distinct}. 
Take the case $J_1/J_2=1.0$ for example, from the inset it is clear that
$\tilde{\chi}_{r}(\mathbf{p})$ is peaked at $\mathbf{p}=(\pi,\pi)$.
Accordingly, the FRG flow for ${\chi}(\pi,\pi)$ (in orange) is most dominant
and rises rapidly as $\Lambda$ is reduced. The flow breaks down around
$\Lambda^*\approx 0.2$, signaling a physical divergence and the onset of
magnetic long-range order as seen in many FRG calculations.  Thus the Neel
phase can be identified unambiguously from the $(\pi,\pi)$ peak and the flow
divergence.

Outside the Neel phase, the flows appear smooth down to the lowest $\Lambda$.
This is perhaps not that surprising because spin rotational symmetry is not
broken in the dimer or plaquette phase.  Yet by inspecting the two cases
$J_1/J_2=0.3$ and $0.7$ in Fig. \ref{fig:model}, it is apparent that their
spin correlations are rather different, e.g., they have different peak momenta
or leading channels. Unfortunately, the information contained in
$\chi(\mathbf{p})$ or $\tilde{\chi}_\alpha(\mathbf{p})$ is too crude to
differentiate the dimer from the plaquette phase. In what follows, we show
that this can be achieved by the FRG flow of bond-resolved spin correlation
$\chi_{ij}$.

\begin{figure}[t]
  \includegraphics[width=0.45\textwidth]{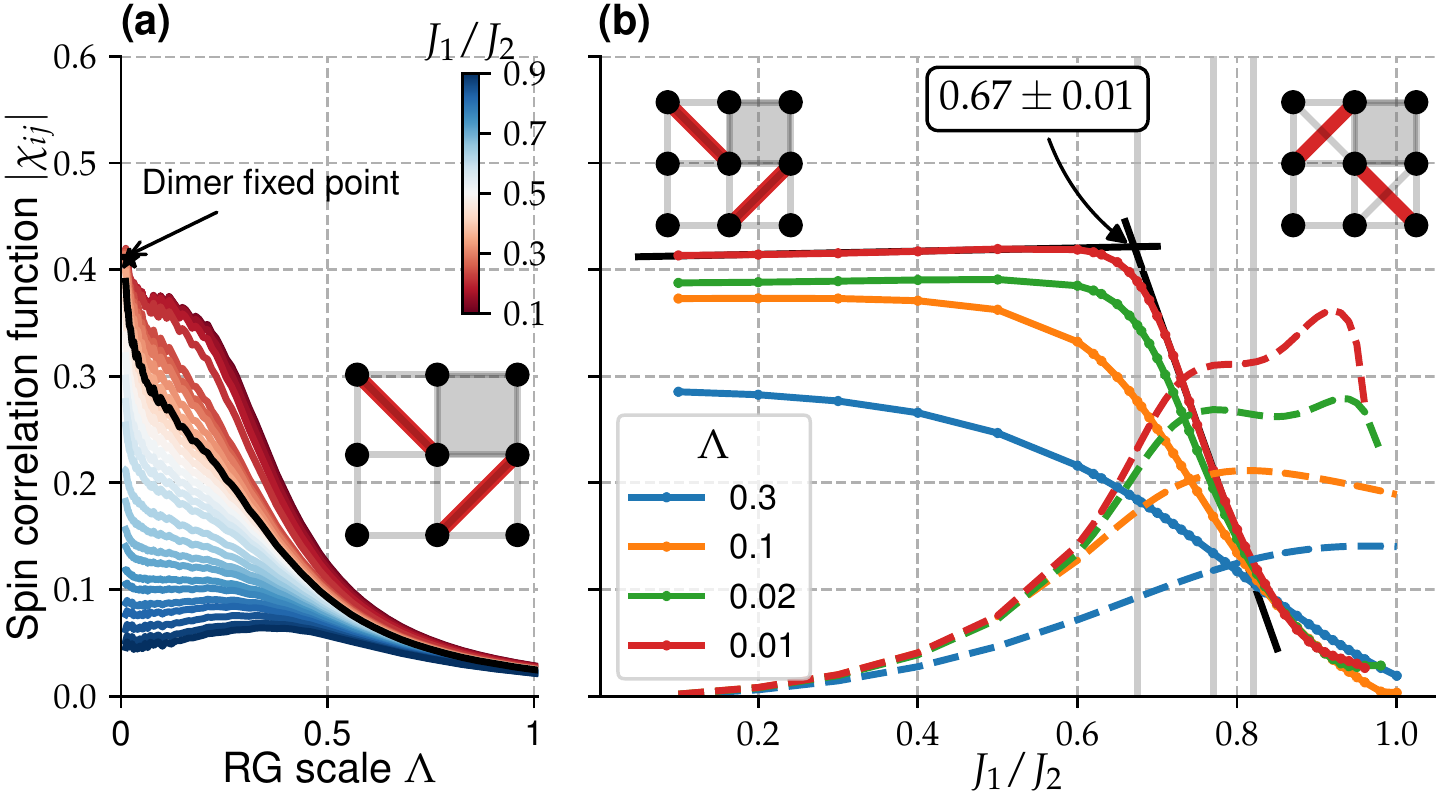}
  \label{fig:susceptibility_real_space}
   \caption{The dimer and spin liquid phase as the fixed points of the flow of spin-spin
   correlation functions $\chi_{ij}(\omega\rightarrow 0)$.
    (a) Flows of $\chi_{ij}$ for the dimer bonds (red lines in inset)
     converge to a constant $\approx 0.41$ in the infrared limit for
     all $J_1/J_2$ values (color coded, see the colorbar) up until 0.67 indicated by the solid black line. 
     (b) The scale-dependent $\chi_{ij}$ (solid lines) for the dimer bonds 
     becomes flat against $J_1/J_2$ in the infrared limit. Linear regression (black lines)
     gives the dimer to plaquette transition point
     $(J_1/J_2)_c=0.67$. The dashed lines represent $\chi_{ij}$ for another set of
     diagonal bonds (inset at the upper right corner) orthogonal to the dimer bonds. 
     As $\Lambda\rightarrow 0$, it becomes flat for $J_1/J_2$ between 0.77 and 0.82
     where spin liquid is postulated to exist.  
     }
\end{figure}


\emph{Dimer phase as a fixed point.---}Fig.~\ref{fig:susceptibility_real_space}(a) compares the flows of $\chi_{ij}$ for
the diagonal bond (red lines in the inset) at different values of $J_1/J_2$.
One notices a remarkable phenomenon: for all $J_1/J_2<0.6$, they flow to the
same exact value $\approx 0.41$ in the infrared $\Lambda\rightarrow 0$.  This
renormalization group fixed point defines a robust phase with constant spin
correlation along the diagonal. This is nothing but the dimer phase, in
accordance with the known fact that the ground state wave function in this
region is a product state of isolated spin singlets, frozen with respect to
$J_1/J_2$ with constant energy up to a critical point.
To determine its phase boundary, 
Fig.~\ref{fig:susceptibility_real_space}(b) plots the diagonal bond correlation
in the infrared limit versus $J_1/J_2$.  It stays completely flat before
dropping rapidly in a linear fashion.  Linear regression (black lines) yields
an intersection point at $J_1/J_2=0.67$ which we take as the estimated phase
transition point.
This critical value is impressively close to 0.675 from large-scale DMRG
\cite{PhysRevX.9.041037}. The agreement provides strong evidence for the
validity and accuracy of our FRG calculation.

\begin{figure}[t]
  \centering
  \includegraphics[width=0.45\textwidth]{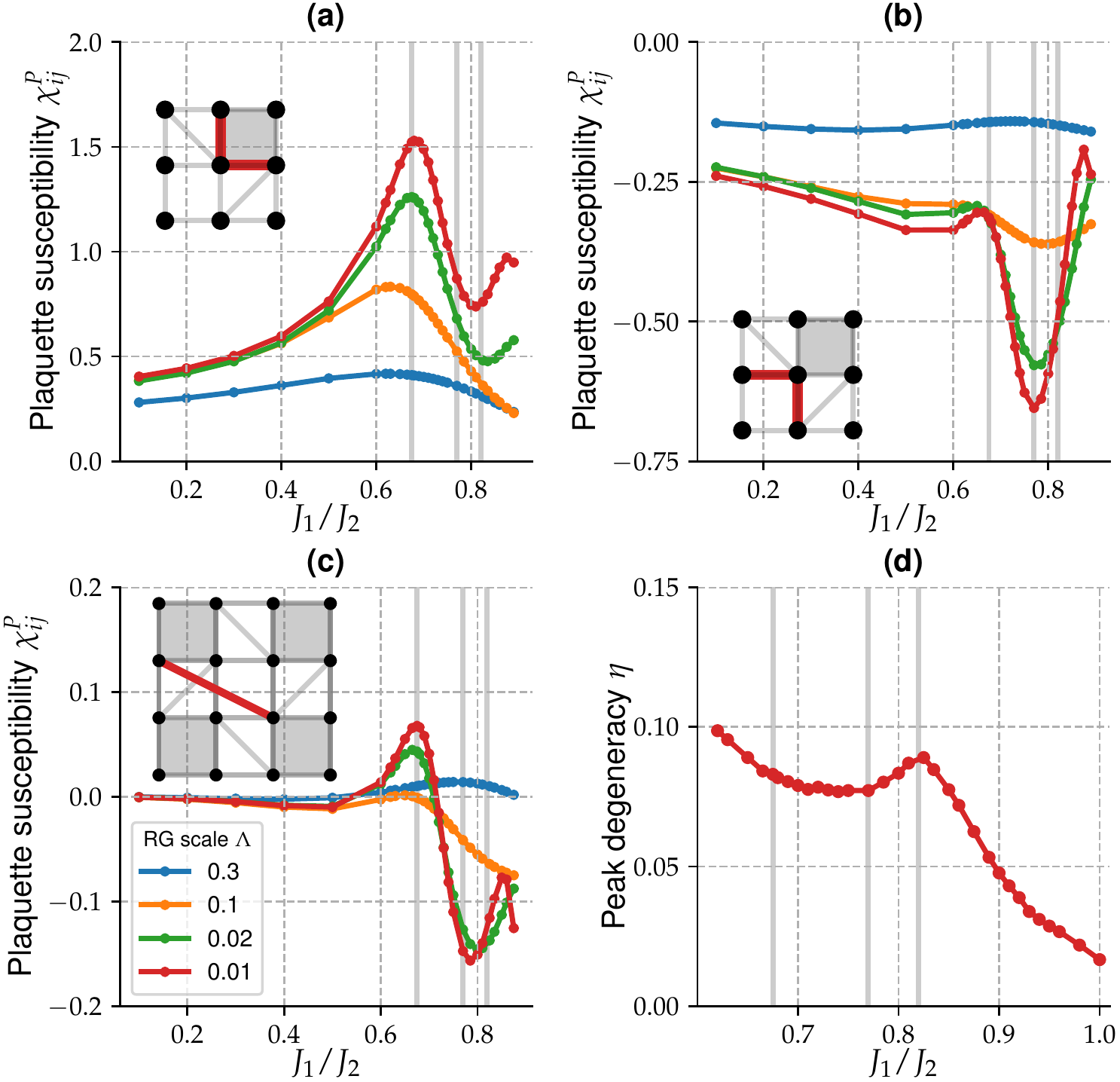}
  \caption{Identification of the phase boundaries from bond-resolved plaquette susceptibility $\chi^P_{ij}$,
  which measures the response of a given bond $(i,j)$ [red lines in insets] to a small increase $\delta J_1$ in the
  shaded squares. 
    (a) A pronounced peak of $\chi^P_\blacksquare$ around $J_1/J_2=0.67$
    signals the onset of plaquette order.
    (b) The suppression dip of $\chi^P_\square$ at $0.77$ indicates the plaquette order
    is superseded by a new phase, the spin liquid with distinct inter-plaquette correlations. 
    (c) An example of longer range $\chi^P_{ij}$ that also develops peak/dip at the two critical points above.
    (d) The momentum space degeneracy $\eta$ of 
    $\tilde\chi_r(\mathbf{p})$
    is peaked at $J_1/J_2\approx 0.82$, which marks
    the transition into the Neel phase. 
       }
  \label{fig:susceptibility_plaquette}
\end{figure}


\emph{Plaquette valence bond solid.---}Identification of plaquette order from
FRG has been an open challenge.  In earlier studies, a plaquette
susceptibility $\chi_{PVB}$ was defined as the propensity towards
translational symmetry breaking with respect to a small bond modulation bias 
\cite{PhysRevB.81.144410,PhysRevB.94.140408,PhysRevX.9.011005}.
Enhancement of $\chi_{PVB}$  has been reported, but to our best knowledge,
plaquette order has not been  positively identified using pseudofermion FRG so
far. For the generalized SS model, we have confirmed that $\chi_{PVB}$ is
indeed enhanced within a broad region stretching from $J_1/J_2\sim 0.5$ to
$0.7$ when compared to its values within the Neel phase (see
\cite{supplementary} for details). But it only exhibits a smooth crossover
with $J_1/J_2$ due to the lack of bond
resolution.  This has motivated us to introduce a more refined measure, the
bond-resolved plaquette susceptibility ${\chi}^P_{ij}$ in Eq.
\eqref{eq:plaquette-susceptibility}. 

Fig.~\ref{fig:susceptibility_plaquette}(a) illustrates the ${\chi}^P_{ij}$ for
the horizontal/vertical bonds within the slightly strengthened plaquettes (shaded squares), 
denoted by $\chi^P_\blacksquare$.
While it is more or less featureless at the ultraviolet scale, as RG steps
are taken and $\Lambda$ is reduced, $\chi^P_\blacksquare$ gains nontrivial dependence on
$J_1/J_2$.  In particular,  in the infrared limit ${\chi}^P_\blacksquare$
develops a pronounced peak around $J_1/J_2\approx 0.67$ \footnote{We have
  checked that the location of the peak remains the same upon increasing the
  correlation cutoff $R_{max}$, adopting a finer frequency grid with lower
  $\Lambda_{IR}$ and taking smaller bias $\delta J_1$.}.  The dramatic
enhancement of plaquette susceptibility marks the onset of plaquette order.
This independent diagnosis of the dimer-to-plaquette transition agrees very
well with the linear regression result above, showing the self-consistency of our FRG
and the advantage of introducing the quantity ${\chi}^P_{ij}$.  It is not a
divergence because higher-order vertices are truncated in the current
implementation. 
The onset of plaquette order also manifests in
longer range inter-plaquette correlations. Fig.~\ref{fig:susceptibility_plaquette}(c) 
depicts the $\chi^P_{ij}$ for a bond between an $r$-site and a $b$-site from two shaded 
squares along the lattice diagonal. It too has an enhancement peak at $J_1/J_2\approx
0.67$.

Further analysis of ${\chi}^P_{ij}$ also points to the demise of the plaquette phase. A pristine 
plaquette order is adiabatically connected to the limit of decoupled plaquette singlets 
(shaded squares in Fig.~\ref{fig:model} without red or dashed bonds). 
Upon increasing $J_1/J_2$, the plaquette order eventually 
yields to a state with homogeneous bond energies and very different spin correlations.
One possibility is a liquid state where the shaded and empty squares
are entangled to feature strong inter-plaquette correlations. The change in correlation is 
apparent in Fig.~\ref{fig:susceptibility_plaquette}(c): after the peak, $\chi^P_{ij}$  
changes sign to develop a sharp dip at $J_1/J_2\approx 0.78$,
suggesting the onset of a new phase. This interpretation is supported by 
the plaquette susceptibility $\chi^P_\square$ shown in  Fig.~\ref{fig:susceptibility_plaquette}(b).
It measures the change to the bonds around the empty plaquettes in response to
$\delta J_1>0$ in the nearby shaded squares. When $J_1/J_2$ is reduced from 
above toward 0.77, a small $\delta J_1$ leads to significant weakening of the 
antiferromagnetic bonds (red lines) around the empty squares, i.e. 
decoupling of the shaded plaquettes to break translational symmetry. 
Thus the pronounce dip of $\chi^P_\square$ at $J_1/J_2\approx 0.77$ 
\footnote{
We note that the suppression dips in Fig.~\ref{fig:susceptibility_plaquette}(b)
and Fig.~\ref{fig:susceptibility_plaquette}(c) do not coincide exactly. 
This discrepancy is expected to reduce with the inclusion of 
higher order vertices.
}
marks the upper critical point of the plaquette phase.
At the very least, the dramatic variations of $\chi^P_{ij}$ are 
at odds with the scenario that the plaquette phase persists after $J_1/J_2\approx 0.77$.


\emph{A sliver of spin liquid.---}The existence of a novel phase after
$J_1/J_2\approx 0.77$ can be inferred independently from the spin correlation
$\chi_{ij}$ for the diagonal bond shown in
Fig.~\ref{fig:susceptibility_real_space}(b) (dashed lines). Here it
becomes flat, i.e. independent of $J_1/J_2$, in the infrared limit.  The
behavior is distinct from that of a plaquette valence bond solid or a Neel
antiferromagnet, for which $\chi_{ij}$ increases with $J_1$.  Since the
spin susceptibility flow is continuous down to $\Lambda\rightarrow 0$ as shown in
Fig.~\ref{fig:model}(b), the only plausible scenario seems to
be that this FRG fixed point corresponds to a liquid phase. 
%
%
With further increase in $J_1$, the flat top of diagonal $\chi_{ij}$ is
terminated by an upturn around $J_1/J_2\sim 0.82$, signaling another phase
transition.  To precisely locate the onset of the Neel order, we adopt an
independent criterion \footnote{The Neel order is rather weak for
  $J_1/J_2<0.9$, so it is numerically hard to pinpoint at what $J_1/J_2$ value
  exactly the divergence sets in. It is more accurate to consider the
  degeneracy of susceptibility as defined in the main text.}.  In the postulated spin liquid region, the
spin susceptibility $\tilde\chi_r(\mathbf{p})$ develops broad maxima, instead of a
sharp peak, in momentum space, see 
the case of $J_1/J_2=0.8$ in Fig.~\ref{fig:model}(b). We can quantify the peak
degeneracy by $\eta$, the percentage of $\mathbf{p}$ points with
$\tilde\chi_r(\mathbf{p})\geq 0.9 \mathrm{max}[\tilde\chi_r(\mathbf{p})]$. A similar
method was employed in \cite{PhysRevX.11.031034} for a different
  system. The result is shown in 
Fig.~\ref{fig:susceptibility_plaquette}(d).  As the Neel phase is
approached, the broad maxima coalesce into a sharp peak at $(\pi,\pi)$, after
which $\eta$ drops quickly. The peak location of degeneracy $\eta$ at $J_1/J_2=0.82$ serves as an
accurate estimation for the transition from the spin liquid to the Neel phase, in
excellent agreement with the phase boundary obtained from large scale DMRG
\cite{yang2021quantum}. 


\emph{Conclusions.---}Our high-resolution FRG analysis of the SS model
identifies four phases separated by three critical points summarized in
Fig.~\ref{fig:model}.  Key technical insights are retrieved by monitoring the renormalization
group flows of bond-resolved spin-spin correlation functions and
susceptibilities.  The good agreement with other established numerical methods
on the locations of the phase boundaries attests to the accuracy of FRG which
takes into account quantum fluctuations in all the channels without bias by
tracking millions of effective couplings at each scale $\Lambda$.  The
implementation and analysis strategies reported here can be applied to study
other quantum spin Hamiltonians with unconventional magnetic orders using
pseudofermion FRG.
In particular, our result supports the existence of a finite spin liquid phase
rather than a deconfined quantum critical point between the plaquette and Neel
phase.  It motivates future theoretical work to further elucidate the nature
and extent of this phase, and precision measurements to locate and probe spin
liquid in $\mathrm{SrCu_2(BO_3)_2}$.

\begin{acknowledgments}
This work is supported by TUBITAK 2236 Co-funded Brain Circulation Scheme 2
 (CoCirculation2) Project No. 120C066 (A.K.) and NSF Grant No. PHY- 2011386 (E.Z.).
\end{acknowledgments}
\bibliography{refs}
\end{document}


\title{Supplementary material for ``Rise and fall of plaquette order in
  Shastry-Sutherland magnet revealed by pseudo-fermion functional
  renormalization group''}
\author{Ahmet Kele\c{s}} 
\affiliation{Department of Physics, Middle East Technical University, Ankara
  06800, Turkey}

\author{Erhai Zhao} 
\affiliation{Department of Physics and Astronomy,
  George Mason University, Fairfax, Virginia 22030,
  USA}
\begin{abstract}  

\end{abstract}
\pacs{} \maketitle
In the supplementary materials, we give technical details about the
implementation of lattice symmetries in the two-particle irreducible vertex.
We present a classical phase diagram of generalized Shastry-Sutherland model
by taking the static limit of the vertex. We give a detailed scan of the phase
diagram and show flow of susceptibilities in the momentum space. We show the
plaquette susceptibility flow as defined in the previous
FRG studies and further motivate our bond-resolved susceptibility calculations.

\section{Lattice symmetries in Shastry-Sutherland model}

In the Shastry-Sutherland model, there are are four lattice sites per unit
cell which requires four different sites with colors $\alpha=r, g, b,
y$ as reference site.  To avoid additional numerical burden stemming from such
sublattice symmetries we consider the $\alpha=r$ site at the origin $i_0$  and
consider a cluster of sites with index $i_2$ within the radius
$|\mathbf{r}_{i_0}-\mathbf{r}_{i_2}|<R_\mathrm{max}$ independent of
sublattice type at $i_2$. In the flow equations and susceptibility
calculations, whenever we need vertex functions of the form
$\Gamma_{j i_2}(s,t,u)$ with a site $j$ of different color $\alpha=g,b,y$,
which is not the
sublattice type of the central reference site, we use the following lattice
symmetry transformations \cite{Reuther2011_1000023236}  
\begin{equation}
    \Gamma_{j i_2}(s,t,u) = \Gamma_{i_{0}j'}(s,t,u) 
    \quad\text{where}\quad j'=
     \begin{cases}
        ( \delta x,  \delta y) &\mbox{ if } j \text{ is } r  \\
        ( \delta y, -\delta x) &\mbox{ if } j \text{ is } g  \\
        (-\delta x, -\delta y) &\mbox{ if } j \text{ is } b  \\
        (-\delta y,  \delta x) &\mbox{ if } j \text{ is } y  
     \end{cases}
     \label{eq:lattice_symmetries}
\end{equation}
where $(\delta x,\delta y)=\mathbf{r}_{i_2}-\mathbf{r}_j=(x_{i_2} -x_j,y_{i_2}-y_{j})$. Furthermore,
the vertex functions have the following symmetries
\begin{eqnarray}
\Gamma_{i_1i_2}( s, t, u) 
&\underset{s\rightarrow -s, i_1\leftrightarrow i_2}{=}& 
 \Gamma_{i_2i_1}(-s, t, u) \nonumber \\
&\underset{t\rightarrow -t}{=}&
 \Gamma_{i_1i_2}( s,-t, u)  \\
&\underset{u\rightarrow -u, i_1\leftrightarrow i_2}{=}&
 \Gamma_{i_2i_1}( s, t,-u) \nonumber \\
&\underset{t\rightarrow -t}{=}&
 \Gamma_{i_1i_2}( u,-t, s)  \nonumber
 \label{eq:frequency_symmetries}
\end{eqnarray}
which is used to set up a frequency grid only for positive $s, t$ and $u$
which reduces the computational cost significantly. Exchange of the site
indices $i_1$ and $i_2$ also has to be done in a way to respect the
sublattice symmetry. In the vertex $\Gamma_{i_1i_2}$, we assume $i_1$ is the
reference site at the origin, $i_2$ is a site with coordinates $(x,y)$ then
the site flipping operation is done by
$\Gamma_{i_1i_2}\rightarrow\Gamma_{i_2i_1}\equiv\Gamma_{i_1,k}$ where $k$ is
the index of the site 
\begin{equation}
    \Gamma_{i_2i_0}(s,t,u) = \Gamma_{i_{0}i_2'}(s,t,u) 
    \quad\text{where}\quad i'_2=
     \begin{cases}
        (-x,-y) &\mbox{ if } i_2 \text{ is }  r  \\
        (-y, x) &\mbox{ if } i_2 \text{ is }  g  \\
        ( x, y) &\mbox{ if } i_2 \text{ is }  b  \\
        ( y,-x) &\mbox{ if } i_2 \text{ is }  y  
     \end{cases}
     \label{eq:inversion_symmetry}
\end{equation}
for $r_{i_2}=(x,y)$ and $r_{i_0}=(0,0)$.

Fourier transform with sublattice degrees
\begin{equation}
    \chi(\omega,\mathbf{k}) = \frac{1}{4} \sum_{i\in\alpha} \sum_j
    e^{i\mathbf{k}\cdot(\mathbf{r}_{i_\alpha}-\mathbf{r}_j)}
    \chi_{ij}(\omega)  
    = \frac{1}{4}\left[
      \chi_r(\omega, k_x, k_y) +
      \chi_r(\omega, k_y,-k_x) +
      \chi_r(\omega,-k_x,-k_y) +
      \chi_r(\omega,-k_y, k_x) 
    \right]  
    \label{eq:susceptibility_definition}
\end{equation}
where we defined the susceptibility contribution from the central site in the
unit cell
\begin{equation}
    \chi_r(\omega,\mathbf{k}) = \sum_j 
    e^{ -i\mathbf{k}\cdot\mathbf{r}_j} \chi_{i_0j} (\omega)  
\end{equation}
and the lattice is setup with $\mathbf{r}_{i_\alpha}=(0,0)$. The last equality
in Eq.~\ref{eq:susceptibility_definition} can be proven using the symmetry
relations given in
Eq.~\ref{eq:lattice_symmetries} and \ref{eq:inversion_symmetry}. 

\begin{figure}
  \centering
  \includegraphics[width=0.30\textwidth]{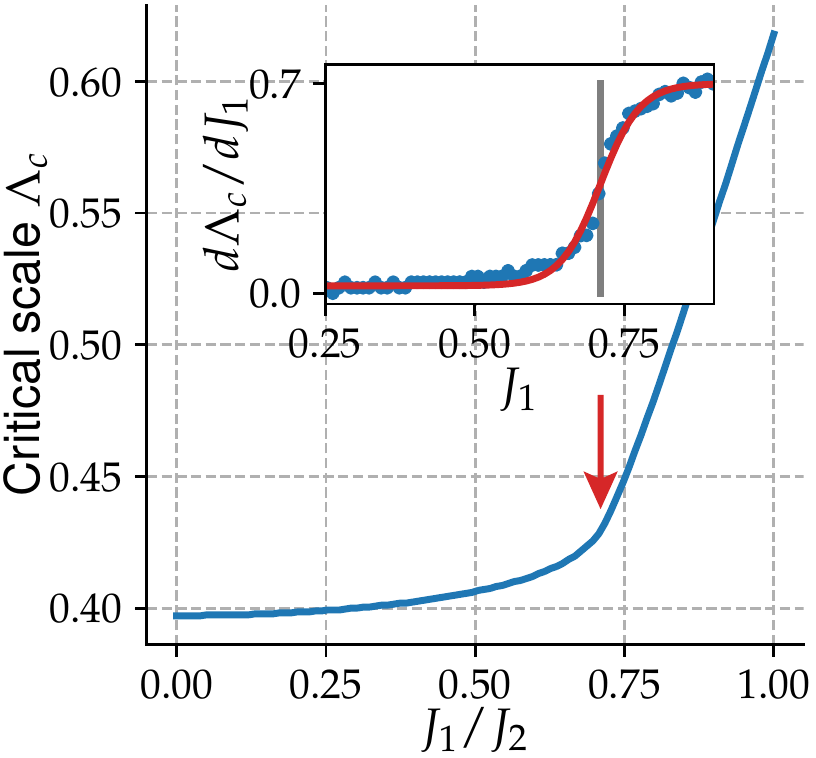}
  \includegraphics[width=0.35\textwidth]{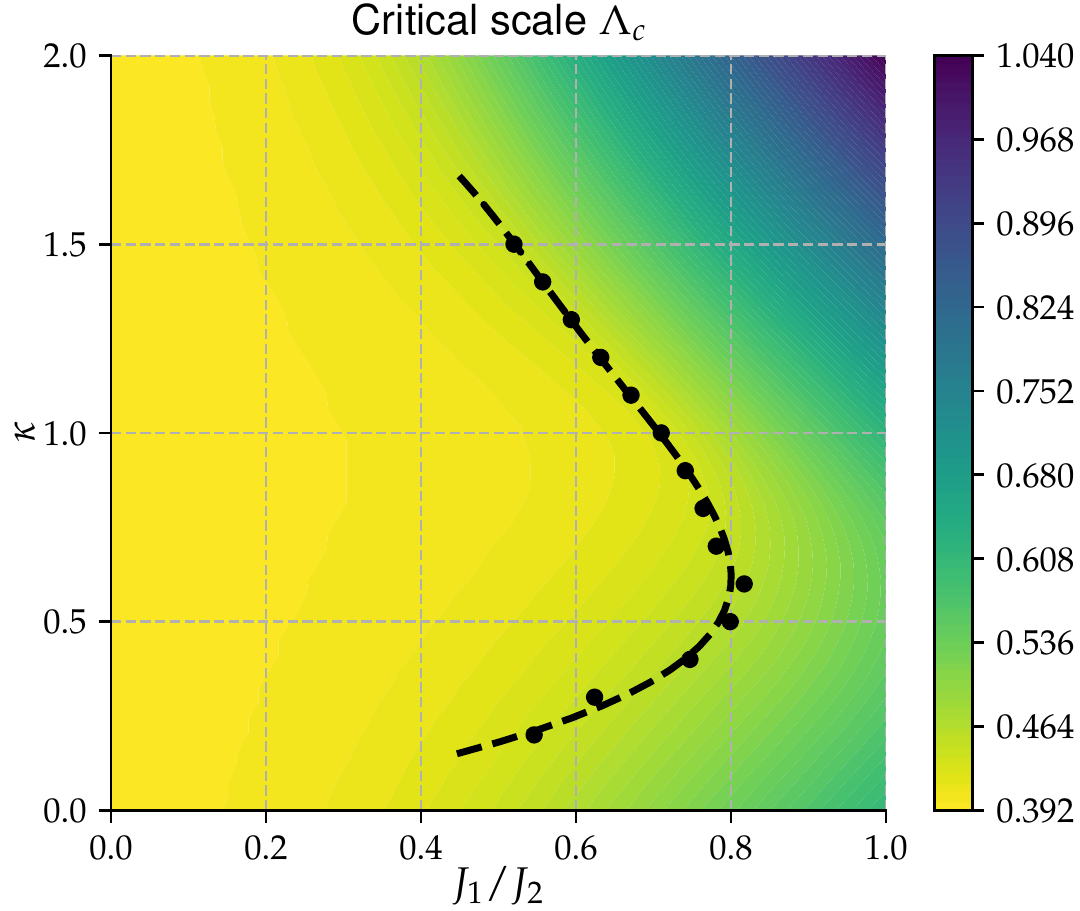}
  \caption{{\bf Left}: Critical scale of Shastry-Sutherland model obtained
    from the static limit of FRG equations for a 5x5 lattice. Critical scale
    is insensitive to $J_1$ on the left side consistent with decoupled
    dimers phase and displays a kink around $J_1/J_2\approx 0.71$ which is
    found from fitting the derivative to a step-like function, see text for
    details. This transition point corresponds to the midpoint of the phase
    transition points $0.675$ and $0.77$ reported in \cite{PhysRevX.9.041037}.
    On the left side of arrow we found divergence in the $\pi-0$ channel
    indicating the dimer phase and on the right we found divergence in
    $\pi-\pi$ channel indicating the Neel order. {\bf Right}: Critical scale
    of the quadrumerized Shastry-Sutherland model obtained from static FRG to
    be compared with the phase diagram in \cite{PhysRevB.66.014401}. Dashed
    black line is a polynomial fit to the data points of the derivatives
    $d\Lambda_c/dJ_1$ for different $\kappa$ shown with filled black circles.}
  \label{fig:static_frg}
\end{figure}
\section{Static vertex limit: Classical phase diagram} 

Calculation of the flow equations for frequency independent vertices gives the
classical phases of the system and corresponds to Random Phase Approximation
(RPA) known in the study of correlated electrons. Here, we include results of
our calculation in static limit, which gives a good starting point for
understanding the phase diagram.

In the limit of static vertex, self energy vanishes and scale dependent
propagator becomes
\begin{equation}
    G^\Lambda(\omega) = 
    \frac{\Theta(|\omega|-\Lambda)}{i\omega+i0^+\sign(\omega)}  
\end{equation}
where we have added an infinitesimal part in the denominator. 
Flow equation for the vertex becomes
\begin{equation}
    \frac{d}{d\Lambda}\Gamma_{si_1i_2} =
    \frac{2}{\pi\Lambda^2}\left[
    \sum_j\Gamma_{si_1j}\Gamma_{sji_2} -2\Gamma_{si_1i_2} ^2
    +\Gamma_{si_1i_2}\Gamma_{si_1i_1}  
    \right] 
    \label{eq:static_frg}
\end{equation}
The first term in
the susceptibility diagram in the main text can be
calculated as
\begin{equation}
    \delta_{ij} 
    \mathrm{Tr}\left[\frac{1}{2} \sigma^z \frac{1}{2} \sigma^z\right] 
    \int_{-\infty}^{\infty}  
    \frac{d\omega}{2\pi} G^\Lambda(\omega) G^\Lambda(\omega)  
    =\delta_{ij}\frac{1}{2} 2\int_\Lambda^\infty
    \frac{d\omega}{2\pi} \frac{1}{(i\omega+i0^+)^2} =
    \delta_{ij}\frac{1}{2\pi(\Lambda+0^+)}  
\end{equation}
The second term with the static vertex can be calculated similarly giving
\begin{equation}
    \chi_{ij}^\Lambda = \frac{1}{2\pi\Lambda} \delta_{ij} -
    \frac{1}{\pi^2\Lambda^2}\Gamma_{sij}^\Lambda 
\end{equation}
In the numerical solution of flow equation Eq.~\ref{eq:static_frg}  in static limit , vertex
functions tend to diverge at small energies signaling
the onset of classical long-range orders. 
In our numerics, when any value in $\Gamma_{ij}$ reaches a maximum value $\Gamma_{max}$,
we stop the flow and identify the corresponding scale as the critical scale
$\Lambda_c$.

\begin{figure}
  \centering
 \includegraphics[width=0.4\textwidth]{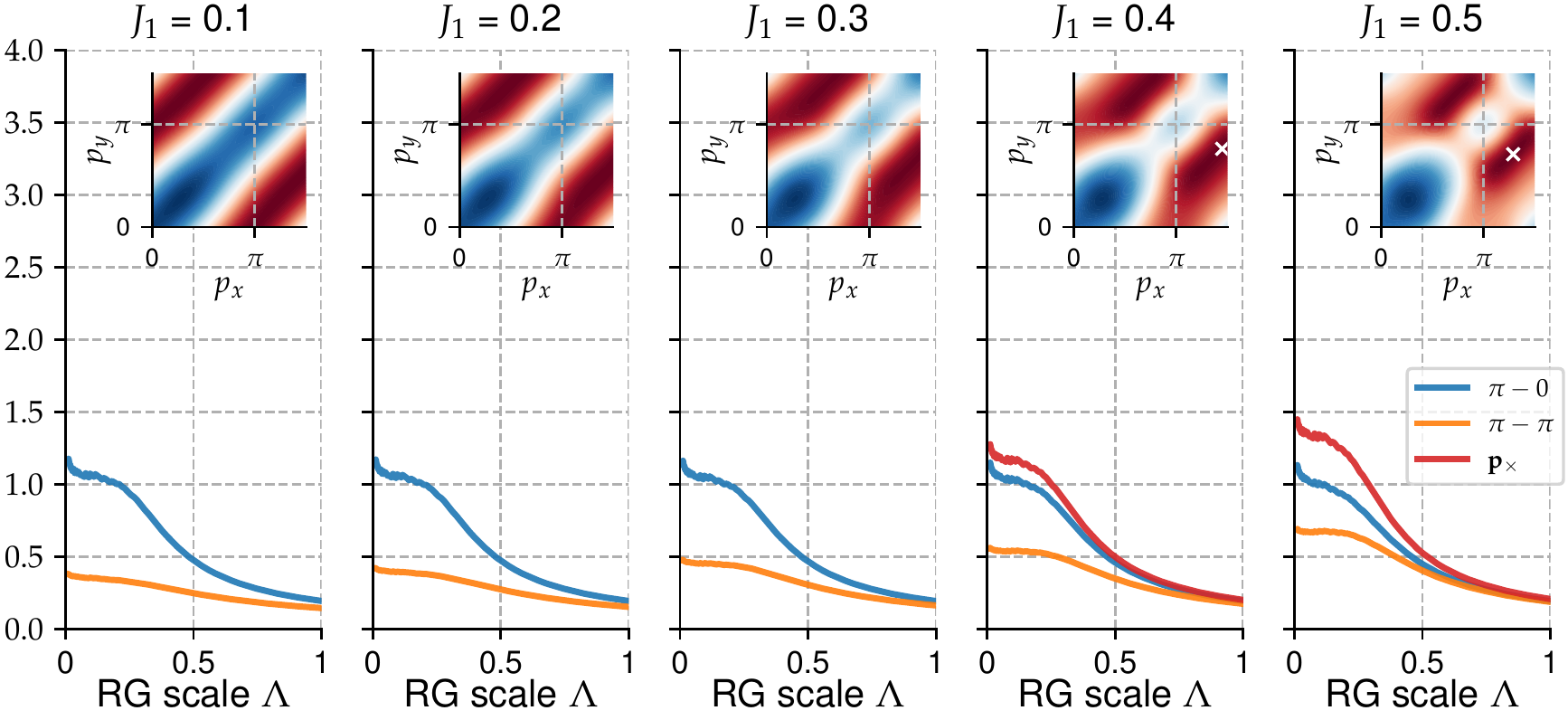}
 \includegraphics[width=0.4\textwidth]{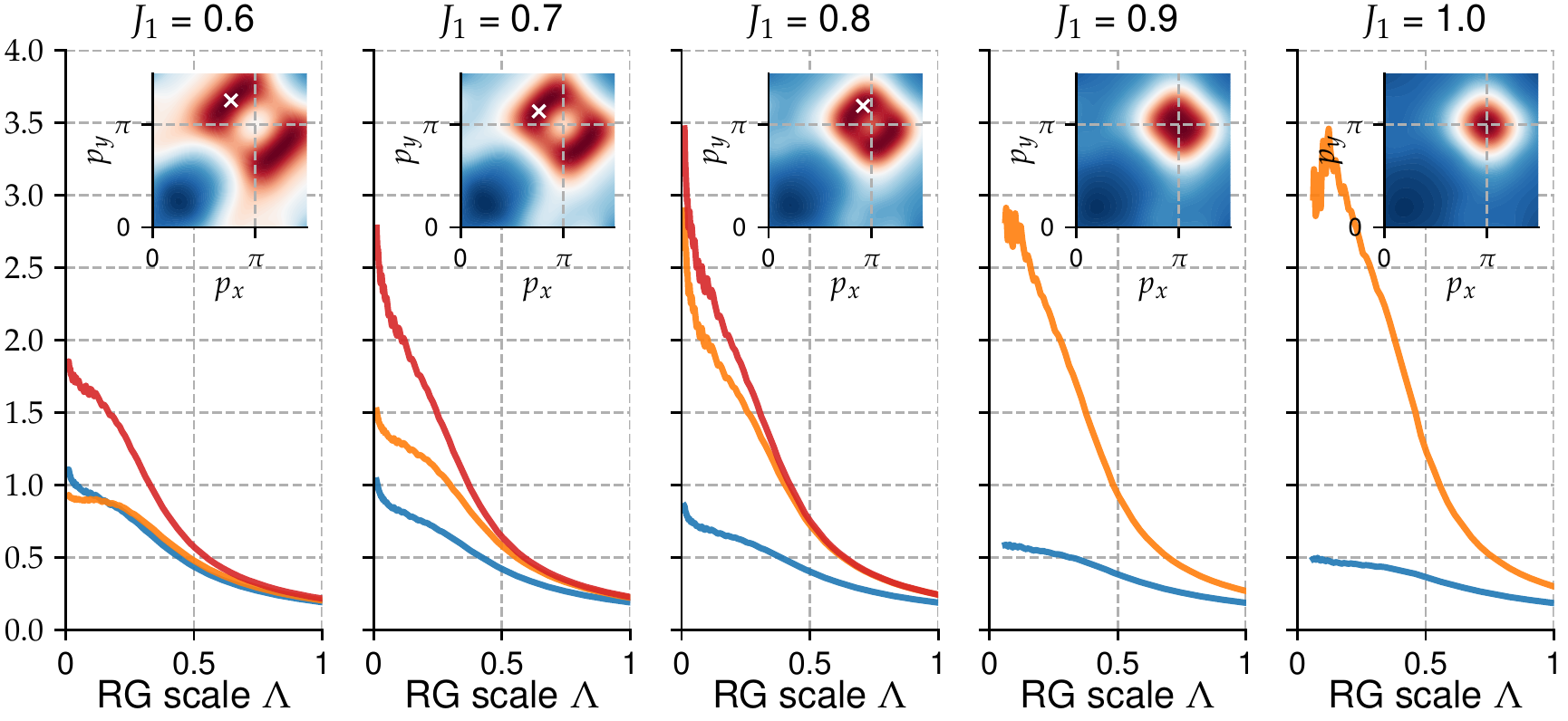}
 \includegraphics[width=0.6\textwidth]{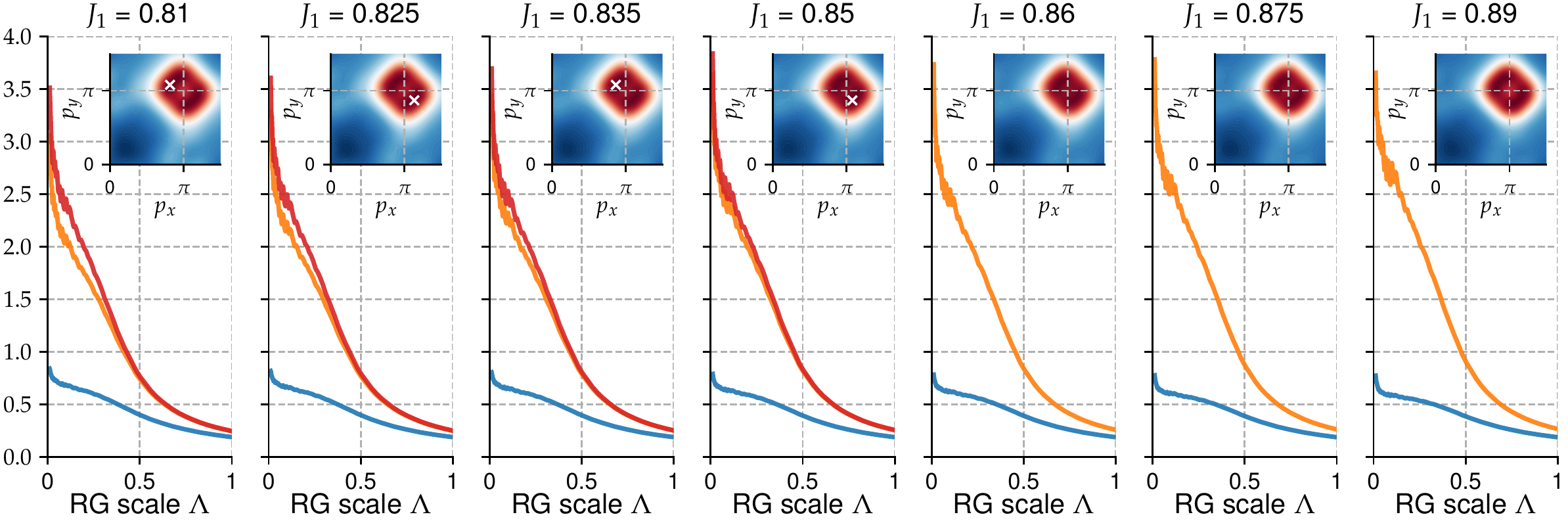}
  \caption{The susceptibility profiles $\chi(\mathbf{p})$.
    Main plot
    shows the flow of maximum susceptibilities and the insets show the
    susceptibility profiles in the upper right corner of the extended
    Brillouin zone at a fixed low RG scale $\Lambda=0.1$ without using
      the sublattice summations in the Fourier transforms. Lower panels
    present a denser scan within the small window between plaquette and Neel
    phases.}
  \label{fig:susceptibility_flows_momentum}
\end{figure}

The critical scale $\Lambda_c$ found from static FRG is shown in
Fig.~\ref{fig:static_frg}. For small $J_1$, the variation of $\Lambda_c$ with
$J_1$ is very small. As we approach the phase transition around $J_1=0.5$,
$\Lambda_c$ start to deviate from this base value somewhat quadratically and
finally it shows
linear increase with $J_1$ inside the known Neel phase. Then, the first
derivative of the transition scale $\Lambda_c$ with respect to $J_1$ is well
approximated with a step-like function. To determine the phase transition
point, we calculate the critical scale $\Lambda_c$ for different
values of $J_1$ and calculate its first derivative $d\Lambda_c/dJ_1$ numerically
and fit the result to a function $a+b\tanh[c(x-x_0)]$. The value for $x_0$
obtained from this fitting with give approximation for the transition
point within the static FRG.  We determine the kink
position $J_c\approx 0.71$, is the
midpoint of the two phase transition points reported in the state-of-the-art
DRMG simulations. This shows the power of pfFRG even in this simplified
calculation. Similar calculation for
generalized SS model with finite $\kappa$ is shown in the right panel of
Fig.~\ref{fig:static_frg}. 

\section{Detailed susceptibility flows in real and momentum space}

Susceptibility flows in momentum space and real space are shown in
Fig.~\ref{fig:susceptibility_flows_momentum} 
with more data
points along the phase diagram.

\section{Plaquette susceptibility without bond resolution} 

We probe the plaquette correlations based on the methodology used in the
previous pfFRG studies\cite{PhysRevB.81.144410,PhysRevX.9.011005} for
comparison. We strengthen the couplings between the spins in the shaded
plaquette (bond towards upper and right nearest neighbors in Fig. 1 in the
main text)
$J_1\rightarrow J_1+\delta J_1$ and weaken the couplings between the spins in
the empty plaquette (lower and left neighbors in Fig.1) $J_1\rightarrow
J_1-\delta J_1$ in the initial conditions by an infinitesimal amount $\delta
J_1$ for fixed $J_2=1$ and calculate the following quantity 
\begin{equation}
    \chi_{PVB} = \frac{J_1}{\delta J_1} 
    \frac{\chi_{ij\in P}-\chi_{ij\notin P}}{\chi_{ij\in P}+\chi_{ij\notin P}}   
\end{equation}
where $\chi_{ij\in P}$ and $\chi_{ij\notin P}$ are susceptibilities of the
bonds in the strengthened and weakened plaquettes, respectively. 
\begin{figure}
  \centering
    \includegraphics[width=0.7\textwidth]{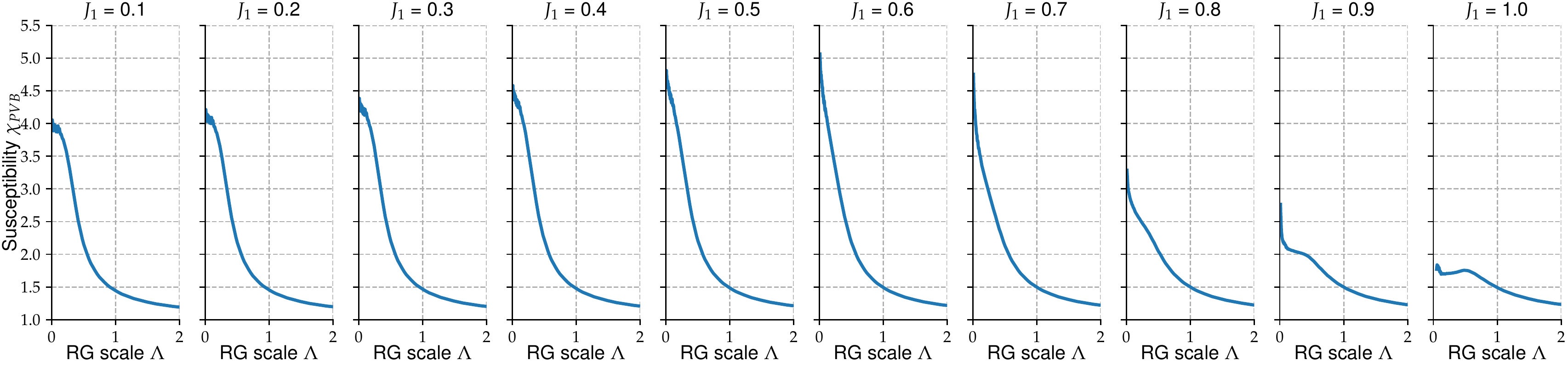}
    \includegraphics[width=0.3\textwidth]{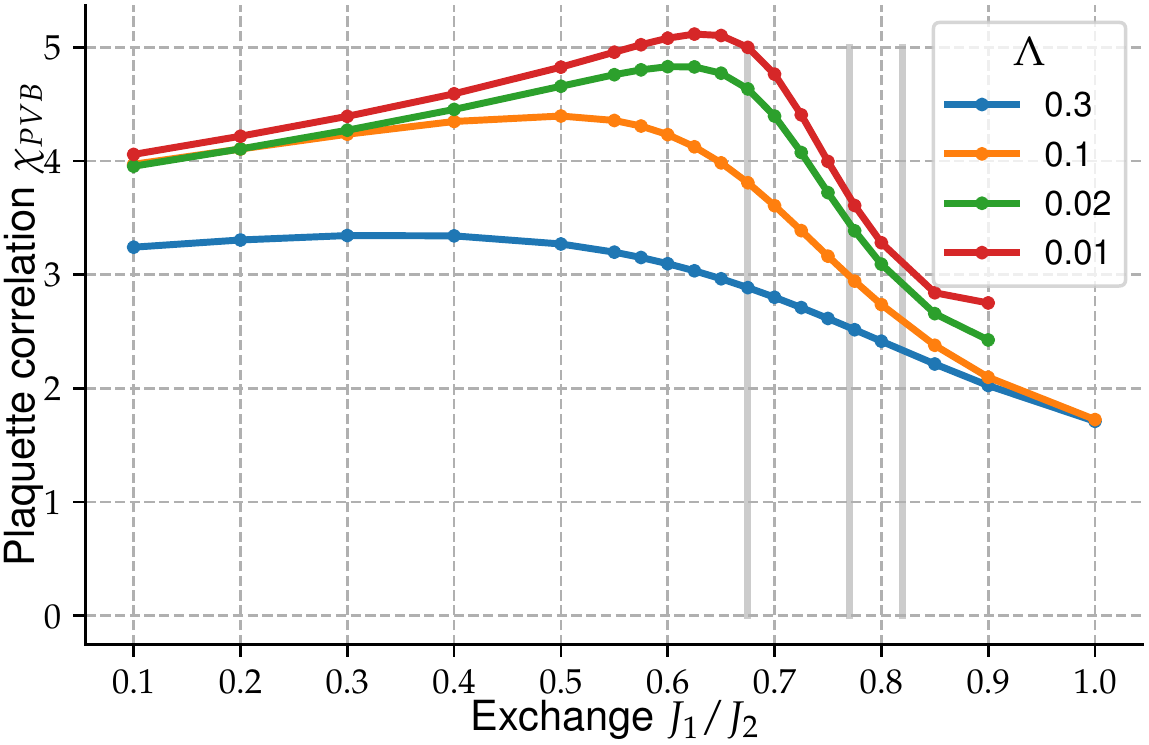}
  \caption{Plaquette susceptibility without bond resolution. Upper panels show
    the flow of susceptibility $\chi_{PVB}$ for selected points along the
    phase diagram. Lower panel shows the denser scan of $\chi_{PVB}$ for three
    different low RG scales.}
  \label{fig:chi_pvb}
\end{figure}

Calculation of this quantity from full FRG is shown in the following
Fig.~\ref{fig:chi_pvb}. Note that there is no bond resolution in
$\chi_{PVB}$.  We observe that it increases about five times around
$J_1=0.6-0.7$, which is in the middle of the putative plaquette phase, and
goes back down as we approach to Neel phase. In the lower panel, we plot
$\chi_{PVB}$
along a denser scan of the phase diagram at different RG scales.

$\chi_{PVB}$ defined above is able to discern short-ranged plaquette phase from
the Neel successfully
but it seems to distinguish dimer and plaquette phases only weakly.  It can be
seen that this susceptibility is significant even for $J_1<0.5$, which was
shown to be the dimer phase unambiguously in the main text.
\section{Definition of bond-resolved plaquette susceptibility} 



One can motivate the alternative bond-resolved susceptibility by considering
the expression $\langle S_i S_j\rangle =
\mathrm{Tr}\left(S_i\cdot S_je^{-\beta H}\right)/\mathrm{Tr}\left(e^{-\beta
    H}\right)$ and taking the derivative of both sides with respect to $J_{kl}$
(coupling between spins $S_k$ and $S_l$) in a way that $i, j, k, l$ are
corners of
a given plaquette pattern while keeping all other couplings
fixed.  It is easy to show $\partial\chi_{ij}/ \partial J_{kl} =
-\beta\langle(S_i\cdot S_j)(S_k\cdot S_l)\rangle+\beta\langle S_i\cdot
S_j\rangle \langle S_k\cdot S_l\rangle$ which is the connected average of the four
spin operators at sites $i, j, k$ and $l$.

To calculate this equation in practice, we start with Shastry-Sutherland model
for a given
nearest-neighbors coupling
$J_1$ by taking $J_2=1$, we calculate the susceptibility $\chi_{ij}[J_1]$ as
usual; we increase
the strength of exchange
coupling  $J_1\rightarrow J_1+\delta J_1$ only for
the target plaquette (upper neighbor and right neighbor couplings $J_1+\delta
J_1$ but
lower neighbor and left neighbor couplings are still $J_1$) and calculate the
resulting susceptibility $\chi_{ij}[J_1+\delta J_1]$; finally define the following
correlation function
\begin{equation}
  \chi^P_{ij} \equiv \frac{\chi_{ij}[J_1+\delta J_1]-
    \chi_{ij}[J_1]}{\delta J_1}.
\end{equation}
This procedure requires running numerically expensive pfFRG
simulation for two points that are infinitesimally close to each other in the
phase diagram but it provides a useful perspective.

\section{Higher-loop contributions}

In this section, we discuss the limitation of our calculation and the possible
extension of our work to include higher-loop contributions. 
To solve the formally exact flow equations, most 
FRG implementations involve two numerical appects, both of which influence the
accuracy and the computational cost of the algorithm. First is the
truncation of the infinite hierarchy of flow equations to a certain loop order
plus higher order correction terms. In the context of pseudo-fermion FRG,
the one-loop truncation plus the Katanin terms as implemented here 
has been extensively applied and benchmarked. For interacting system,
no numerical implementation of FRG manages to include all diagrams to infinite loop order. 
Empirically, most known divergences (instabilities toward long-range order), and the 
mutual influence of these instabilities,
seem to be accounted already at the level of one-loop diagrams for interacting fermions.
Second is the resolution for momentum and frequency
in the numerical solution of the truncated flow equations. The term
``high-resolution" throughout our manuscript refers to the frequency as well as
the momentum resolution (or equivalently, the site resolution, which in our
implementation is determined by the cut-off correlation length in real space).  
A main technical effort of our work is to achieve sufficient frequency and site
resolution, which as we have seen in the main text is crucial to analyze the 
plaquette susceptibility, leading to the identification of 
an unexpected spin liquid region. For example, with 64 grid points in the frequency and 21$\times$21=441 lattice sites, we
end up with 120 million running couplings. Decreasing the frequency resolution from 64 to 48
decreases to the number of coupling to about 50 million but the change in the
flow of spin susceptibilities are almost the same at the level of machine precision.
The computation is made
feasible by migrating to GPU. The improved resolution not only can help identify 
complex long-range orders hitherto
impossible, but also can provide fresh physical insight by comparing the various
bond-resolved susceptibilities.

A natural question is then ``how important are additional higher loop contributions?"
 A definitive answer to this questions would require
solving the flow equation with additional higher-loop terms systematically taken into
account, which is beyond the computing resources available to us at the
moment. But in principle, bond-resolved spin and plaquette susceptibilities 
can be calculated with additional higher-loop contributions. This
would be the subject of a future research. 

Very recently, a few groups have begun to shed light on this open question
by explicitly taking into account additional higher-loop 
contributions \cite{kiese2021multiloop,thoenniss2020multiloop}.
In assessing these results and discussing higher-loop expansions, one must 
carefully distinguish FRG for itinerant fermions and pf-fRG for
quantum spin systems, because the nature of the loop expansion, hence the role
played by the higher loops are drastically different. 
 For itinerant fermions, such as the one-band Fermi-Hubbard model on square
lattice studied by Hille et al. \cite{PhysRevResearch.2.033372}, one
motivation to include higher loop contributions is to accurately describe the
momentum and frequency dependence of the self-energy, which was neglected in
many of the earlier N-patch FRG studies. The other motivation is to achieve
quantitative agreement with Parquet approximation and Determinant quantum
Monte Carlo. It remains to be seen how such ``improved fRG*'' performs 
when away from half-filling.

For quantum spin systems, 
two recent works investigated the higher-loop
corrections in pf-fRG \cite{kiese2021multiloop,thoenniss2020multiloop}. As we outlined in the main text,
the pseudo-fermion representation of
spin models leads to a strongly interacting system where the bare kinetic energy is
zero. Here the loop expansion is motivated more by the spirit of 1/S or 1/N
expansion \cite{PhysRevB.96.045144,PhysRevB.97.064415}.
(In pseudo-fermion FRG, the frequency dependence of self-energy plays a crucial role and
is already taken into account from the very beginning.) 
To our knowledge, there still lacks a satisfactory understanding of the role
of higher-loop contributions, including its convergence properties when 
long-range or spin liquid phases are approached in the infrared limit.
The state-of-the-art multiple-loop pf-fRG seems to imply that
the divergence towards long-range order, or the lack thereof, is dictated by the leading
contributions contained in the level-one truncation (one loop plus Katanin
terms, as implemented in our work). 
Moreover, our confidence in the performance of pf-fRG applied to the
Shastry-Sutherland (SS) model stems from its (unexpected) remarkable
agreement with other independent, state-of-the-art methods such as DRMG
regarding the phase boundaries. Future work will elucidate whether or how
the predicted phase boundary may change with higher order contributions included
in a consistent manner.

\bibliography{refs}
